\documentclass[aip,apl,reprint]{revtex4-1}

\usepackage{graphicx}
\usepackage{dcolumn}
\usepackage{bm}

\begin{document}


\title{Enhanced Transmission in Rolled-up Hyperlenses utilizing Fabry-Pe\'rot Resonances}

\affiliation{Institut f\"ur Angewandte Physik und
Mikrostrukturforschungszentrum, Universit\"at Hamburg,
Jungiusstrasse 11, D-20355 Hamburg, Germany}

\author{Jochen Kerbst} 
\author{Stephan Schwaiger} \email{sschwaig@physnet.uni-hamburg.de}
\author{Andreas Rottler} 
\author{Aune Koitm\"ae} 
\author{Markus Br\"oll} 
\author{Andrea Stemmann} 
\author{Christian Heyn} 
\author{Detlef Heitmann} 
\author{Stefan Mendach}

\date{\today}

\begin{abstract}
We experimentally demonstrate that the transmission though rolled-up metal/semiconductor hyperlenses can be enhanced at desired frequencies utilizing Fabry-P\'erot resonances. By means of finite difference time domain simulations we prove that hyperlensing occurs at frequencies of high transmission. 
\end{abstract}

\maketitle

Metamaterials offer the possibility to tailor their effective optical parameters \cite{Pendry1999, Smith2000} allowing e.g. sub-wavelength imaging. Beside metamaterials with negative index of refraction \cite{Veselago1968, Pendry2000, Fang2005}, metamaterials consisting of alternating layers of metal and dielectric \cite{Anantha2003,Wood2006} are able to transmit electromagnetic waves containing sub-wavelength details of an object. In the latter case sub-wavelength imaging relies on the anisotropic permittivity of the multilayer structure leading to unidirectional propagation of electromagnetic waves. Magnification of a sub-wavelength object can be achieved if the multilayers are curved \cite{Jacob2006}. As a consequence the light is channeled in radial direction. These so called hyperlenses have been fabricated consisting of multilayers of Ag and oxides with an effective plasma frequency in the violet and ultraviolet regime \cite{Liu2007,Rho2010}. They are most efficient close to the effective plasma frequency in the longitudinal component of the permittivity of the metamaterial \cite{Anantha2003}, which can be tuned by the ratio of the thicknesses of the dielectric and metal layer. Recently we have shown that using the concept of self-rolling strained layers \cite{Prinz2000, Schmidt2001, Schumacher2005, Mendach2005a}, one can fabricate a rolled-up metal/semiconductor microtube which exhibits optical anisotropy and acts as a hyperlens \cite{Schwaiger2009}. The operation frequency of these rolled-up hyperlenses (RHLs) is tunable in the visible and near-infrared regime.

In this letter we propose a concept to optimize the transmission through a metamaterial. We utilize Fabry-P\'erot resonances related to the total thickness of the metamaterial to obtain increased transmission in a desired frequency regime. We illustrate this concept for the example of a RHL \cite{Schwaiger2009,Schwaiger2011,Smith2009,Smith2010} consisting of alternating layers of Ag and semiconductor as sketched in Fig. 1(a). We show that the number of rotations of our RHL can be used to maximize the transmission at their operation frequency and obtain values as high as 52$\,\%$. Finite difference time domain (FDTD) simulations prove the hyperlensing ability of the presented structures.

\begin{figure}
\includegraphics{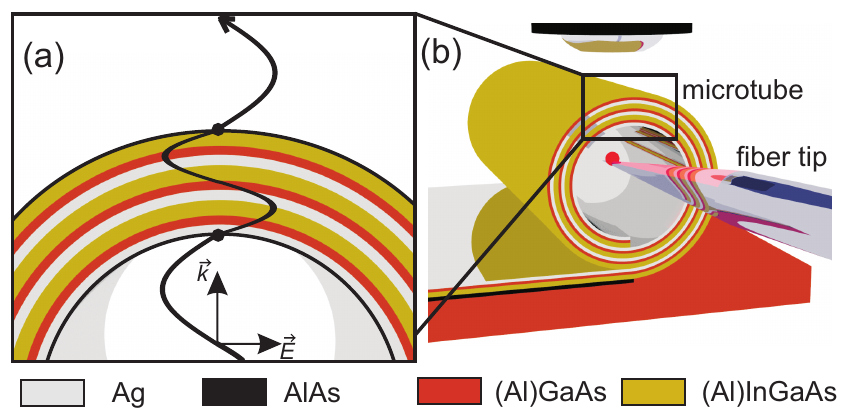}
\caption{\label{fig:tube} (a)~The RHL consists of several alternating layers of semiconductors and Ag. Light is transmitted through the wall of the RHL. The Fabry-P\'erot resonance determined by the total thickness of the wall can be tuned to a desired frequency. (b)~A tapered optical fiber which emits light through a hole in the sidewall perpendicular to the fiber's orientation is manipulated into a microtube. The transmitted light can be detected with a microscope setup.}
\end{figure}

The preparation of the RHLs is described in the following: Initially a semiconductor heterostructure is grown on a GaAs substrate using molecular beam epitaxy. The heterostructure consists of a GaAs buffer layer ($500$~nm), an AlAs sacrificial layer ($40$~nm) followed by a strained Al$_{20}$In$_{13}$Ga$_{67}$As layer ($23$~nm), a strained In$_{16}$Ga$_{84}$As layer ($7$~nm), and an unstrained Al$_{23}$Ga$_{77}$As layer ($21$~nm). On top of these semiconductors an Ag layer is deposited by thermal evaporation. Finally the AlAs sacrificial layer is removed by selective etching and the AlInGaAs/InGaAs/AlGaAs/Ag system minimizes its strain energy by rolling up into a RHL with several rotations.

To perform transmission measurements through the RHLs we use a fiber based transmission measurement setup which is sketched in Fig. 1(b). A tapered single mode fiber with a tip diameter of $d_{\mathrm{tip}}=2\,\mu$m can be manipulated into the microtubes with the help of an XYZ piezo stage. Monochromatic light, which is coupled into the optical fiber is guided to a hole in the sidewall near the tip and emitted perpendicularly to the fiber axis. After being transmitted through the wall of the microtube the light is collected with a near-infrared corrected objective, and detected by an InGaAs photodiode using lock-in technique. For normalization of the spectra we acquire spectra with the fiber tip inside the microtube and spectra with the tip outside the microtube.

\begin{figure}
\includegraphics{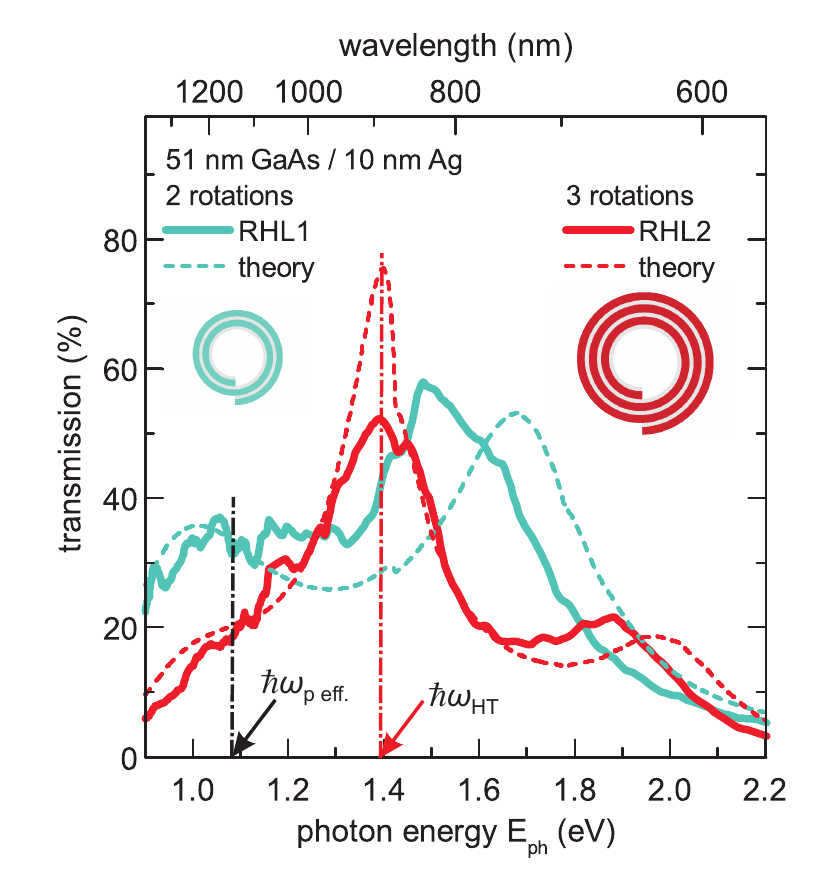}
\caption{\label{fig:Windungszahl} Measured transmission spectra (solid lines) through two RHLs and the corresponding transfer matrix calculations (dashed lines). RHL1 (blue line) consists of two alternating layers of Ag ($d_{\mathrm{Ag}}=10\,$nm) and (AlIn)GaAs ($d_{\mathrm{SC}}=51\,$nm), while RHL2 (red line) exhibits three alternating layers with the same single layer thicknesses as RHL1. The peak at $E_{\mathrm{ph}}=1.39$~eV ($E_{\mathrm{ph}}=1.48$~eV) with $T=52\,\%$ ($T=58\,\%$)  for RHL2 (RHL1) can be attributed to a Fabry-P\'erot resonance in the total thickness of the wall of the microtube. The high transmission frequency of a RHL $\omega_{\mathrm{HT}}$ (red arrow) has to be tuned to the vicinity of the plasma frequency $\omega_{\mathrm{p\,eff.}}$ (black arrow) to assure unidirectional propagation of electromagnetic waves.}
\end{figure} 

In Fig. 2 we compare the transmission through two exemplary hyperlenses RHL1 and RHL2 with identical operation frequency, i.e. identical individual layer thicknesses ($d_{\mathrm{Ag}}~=~10\,$nm) and (AlIn)GaAs ($d_{\mathrm{SC}}~=~51\,$nm), but varying number of rotations. RHL1 exhibits a total thickness of $t_{\mathrm{RHL1}}~=~122\,$nm, while RHL2 exhibits a total thickness of $t_{\mathrm{RHL2}}~=~183\,$nm. In the range from $E_{\mathrm{ph}}~=~1.26$~eV to $E_{\mathrm{ph}}=1.45$~eV the transmission through RHL2 exceeds that one through RHL1 even though the total wall thickness is increased, i.e. $T_{\mathrm{RHL2}}~>~T_{\mathrm{RHL1}}$. The maximum transmission through RHL1 occurs at $E_{\mathrm{ph}}=1.48$~eV with a value of $T=58\,\%$. In the case of the RHL2 the maximum transmission occurs at $E_{\mathrm{ph}}=1.39$~eV with a value of $T=52\,\%$. 

To model the measurements we calculate the transmission using the transfer matrix method \cite{Born1980} and approximate the RHL as a flat superlattice of Ag and GaAs. The dielectric functions of Ag and GaAs were taken from Ref. \cite{Palik1985}. The calculated transmission curves for the fabricated RHLs are presented in Fig. 2 (dashed lines). They show a good agreement with the measured data. To understand the origin of the transmission maxima we perform further calculations. We vary the single layer thicknesses of Ag and GaAs while keeping the layer thickness ratio constant. This allows us to vary the total thickness of the RHL while maintaining the effective optical properties and the effective plasma frequency $\omega_{\mathrm{p\,eff.}}$. We find that the maximum occurring in the calculations on RHL2 can be shifted to lower photon energies with increasing total thickness of the RHL. Furthermore, we double and triple the total thickness ($2\,t_{\mathrm{RHL2}}=366\,$nm and $3\,t_{\mathrm{RHL2}}=549\,$nm) and determine a transmission maximum at the same energetical position $E_{\mathrm{ph}}=1.39$~eV. This clearly shows that the observed transmission maximum is a Fabry-P\'erot resonance related to the total thickness of the RHL.

\begin{figure}
\includegraphics{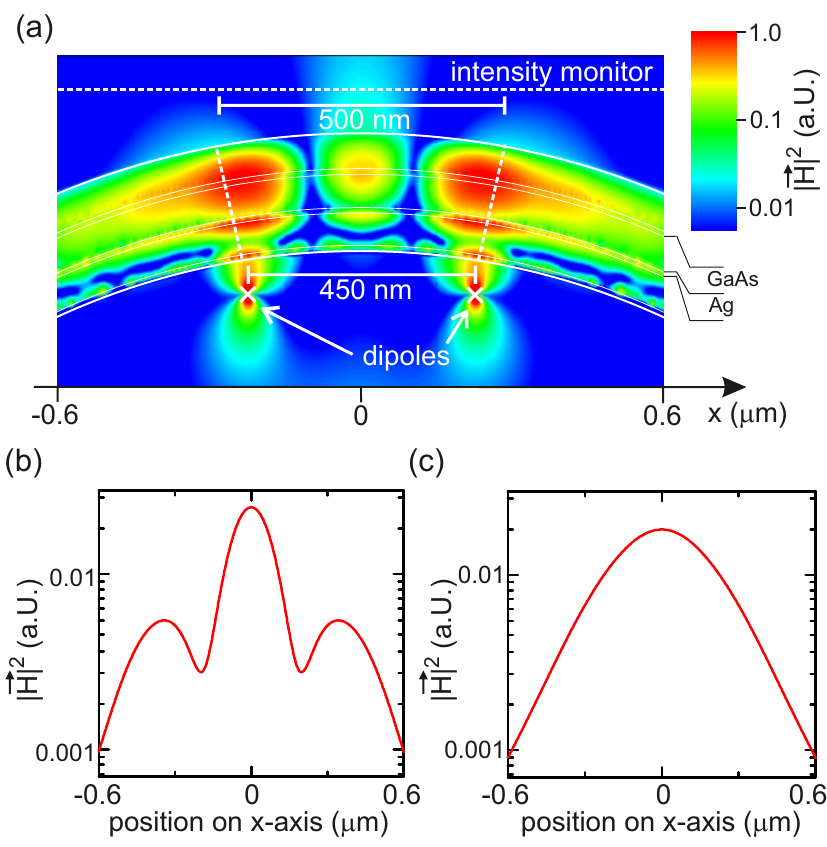}
\caption{\label{fig:Simulation} Finite-difference time-domain simulations (\textit{Lumerical FDTD-solutions}) on RHL2. Two dipoles emitting at $\hbar \omega_{\mathrm{dipole}}=1.39\;$eV are placed at the inner perimeter of RHL2. (a) Logarithmic plot of the magnetic field intensity $|\vec{H}|^2$ in RHL2. (b) Spatial distribution of $|\vec{H}|^2$ in a distance of $0.3\;\mu$m from the dipoles with RHL2 in between. The two dipoles can be resolved. (c) Spatial distribution of $|\vec{H}|^2$ in a distance of $0.3\;\mu$m from the dipoles without RHL2.}
\end{figure}

In order to obtain hyperlensing, the operation frequency of the RHL has to be in the vicinity of the plasma frequency of the effective medium $\omega_{\mathrm{p\,eff.}}$ \cite{Schwaiger2009} which is marked in Fig.~2 with a black arrow. On the other hand the transmission through any metallic material drops below the plasma frequency. A compromise of high transmission and hyperlensing operation quality has to be found in the case of hyperlensing devices. In the following we use FDTD to show that RHL2 exhibits pronounced hyperlensing not only directly at the plasma frequency $\omega_{\mathrm{p\,eff}}$ but also at a nearby Fabry-P\'erot maximum at $\omega_{\mathrm{HT2}}>\omega_{\mathrm{p\,eff.}}$ (Fig. 2 red arrow) where the transmission is increased to $T=52\,\%$. We simulate the transmission through RHL2 using the same optical parameters as in the transfer matrix calculations in Fig~\ref{fig:Windungszahl} and the inner radius corresponding to RHL2 ($r_{\mathrm{in}}\approx2\,\mu$m).

We place two dipoles at a distance of $45\;$nm from the inner surface of RHL2. Both emit at an energy of $E_{\mathrm{dipole}}=1.39\;$eV which corresponds to the maximum transmission of calculations on RHL2. The magnetic field vector $\vec{H}$ was chosen to point along the rolling axis of RHL2 and the electric field vector $\vec{E}$ perpendicular to it. In Fig.~3(a) we show the intensity of magnetic field $|\vec{H}|^2$ in a color plot. We observe that the field is radially channeled which indicates that the structure exhibits an unidirectional propagation of electromagnetic waves. As a consequence, the fields of the two dipoles can be clearly resolved at the outer perimeter. This is also demonstrated in Fig.~3(b) and (c) where we collected the intensity of the magnetic field $|\vec{H}|$ in a distance of $0.3\;\mu$m from the dipoles, with (Fig.~3(b)) and without (Fig.~3(c)) RHL2. Without RHL2 between the sources and the monitor the field of the two dipoles have already merged to a Gaussian-shaped intensity distribution. With RHL2 the peak fields of the two dipoles are still distinguishable, which demonstrates hyperlensing.

In conclusion we experimentally showed that the transmission through a RHL can be optimized by tuning the Fabry-P\'erot resonance related to multiple reflection at the inner and outer surface of the RHL and achieve a transmission value as high as $T=52\,\%$. Furthermore, using FDTD simualtions we demonstrated that hyperlensing is possible at the frequency of high transmission.

We gratefully acknowledge support of the Deutsche Forschungsgemeinschaft via the Graduiertenkolleg 1286 ``Functional Metal-Semiconductor Hybrid Systems'' and LExI ``Nano-Spintronics''.

\appendix

\newpage

\end{document}